\documentclass[conference]{IEEEtran}
\IEEEoverridecommandlockouts
\ifCLASSINFOpdf
  \usepackage[pdftex]{graphicx}
   \graphicspath{{Figures/}}
   \DeclareGraphicsExtensions{.pdf,.jpeg,.png}
\else
\fi
\usepackage{amsmath}
\usepackage{amssymb}
\usepackage{amsthm}
\usepackage{caption}
\usepackage{algorithm}
\usepackage{algorithmicx}
\usepackage{algpseudocode}

\usepackage{array}

\ifCLASSOPTIONcompsoc
  \usepackage[caption=false,font=normalsize,labelfont=sf,textfont=sf]{subfig}
\else
  \usepackage[caption=false,font=footnotesize]{subfig}
\fi
\usepackage{color}
\usepackage{comment}

\newtheorem{theorem}{Theorem}

\newtheorem{definition}{Definition}

\begin{document}
%
\title{Online Influence Maximization in Non-Stationary Social Networks}



%
\author{\IEEEauthorblockN{Yixin Bao\IEEEauthorrefmark{1},
Xiaoke Wang\IEEEauthorrefmark{1},
Zhi Wang\IEEEauthorrefmark{2},
Chuan Wu\IEEEauthorrefmark{1},
Francis C.M. Lau\IEEEauthorrefmark{1}\thanks{The project was supported in part by grants from Hong Kong RGC under the contracts HKU 717812E, C7036-15G (CRF), and the National Natural Science Foundation of China (NSFC) under Grant No.~61402247}}
\IEEEauthorblockA{\IEEEauthorrefmark{1}Department of Computer Science, The University of Hong Kong, Email: \{yxbao,xkwang,cwu,fcmlau\}@cs.hku.hk}
\IEEEauthorblockA{\IEEEauthorrefmark{2}Graduate School at Shenzhen, Tsinghua University, Email: wangzhi@sz.tsinghua.edu.cn}}


\maketitle

\begin{abstract}
Social networks have been popular platforms for information propagation. 
An important use case is viral marketing: given a promotion budget, an advertiser can choose some influential users as the seed set and provide them free or discounted sample products; in this way, the advertiser hopes to increase the popularity of the product in the users' friend circles by the world-of-mouth effect, and thus maximizes the number of users that information of the production can reach. There has been a body of literature studying the influence maximization problem. Nevertheless, the existing studies mostly investigate the problem on a one-off basis, assuming fixed known influence probabilities among users, or the knowledge of the exact social network topology. In practice, the social network topology and the influence probabilities are typically unknown to the advertiser, which can be varying over time, i.e., in cases of newly established, strengthened or weakened social ties. In this paper, we focus on a dynamic non-stationary social network and design a randomized algorithm, RSB, based on multi-armed bandit optimization, to maximize influence propagation over time. The algorithm produces a sequence of online decisions and calibrates its explore-exploit strategy utilizing outcomes of previous decisions. It is rigorously proven to achieve an upper-bounded regret in reward and applicable to large-scale social networks. Practical effectiveness of the algorithm is evaluated using both synthetic and real-world datasets, which demonstrates that our algorithm outperforms previous stationary methods under non-stationary conditions.
\end{abstract}


%
\IEEEpeerreviewmaketitle

\section{Introduction}
Influence maximization in social networks is an important problem that seeks the best seed users to maximize the spread of information \cite{kempe2003maximizing}. Prominent use cases include advertising and viral marketing \cite{kempe2003maximizing}\cite{richardson2002mining}. When a company is promoting a new product, it can engage some influential users as seeds in a social network, providing them samples for free or at discounted prices. These seed users may inform their friends of this product, and their friends will further influence other users, and so on. Through world-of-mouth distribution, the product will get to be known by more and more users in the social network. As it is common for a company to have a promotion budget, it is most beneficial to identify the best set of seeds so as to maximize the number of users that information can eventually reach. 


The influence maximization problem has been studied on several probabilistic cascade models. In the independent cascade model \cite{kempe2003maximizing},
each node probabilistically actives (influences) its neighbors at each time stamp independently of the history thus far, and a node only attempts to activate a neighbor once. 
In the linear threshold model \cite{kempe2003maximizing}, a node will be activated only when the sum of influence probabilities from its neighbors exceeds a threshold. 
The influence probability in the above models, namely the probability for node $u$ to activate its neighbor $v$ after $u$ has been activated, is often decided empirically in studies designing influence maximization algorithms, 
 {\em e.g.}, according to inverse of the indegree of $v$.

Based on these information propagation models, existing studies mostly tackle the influence maximization problem on a one-off basis, assuming that both the social network topology and influence probabilities are fixed and available as input. Kempe {\em et al.}~\cite{kempe2003maximizing} prove that the influence maximization problem is NP hard but can be approximated to within a factor of $(1-\frac{1}{e}-\varepsilon)$ with a greedy hill-climbing method, where $\varepsilon$ is any positive real number. 
A number of other approximation algorithms have also been proposed to achieve near-optimal time complexity, including CELF \cite{leskovec2007cost}, CELF++ \cite{goyal2011celf++}, TIM \cite{tang2014influence}, and IMM \cite{tang2015influence}. In real-world social networks, exact network topology and influence probabilities are typically unknown to a third party advertiser, and are time-varying. For example, new social ties are set up when people make new friends, and the ties can be strengthened over time when they become more familiar; two people become connected when collaborating on a short-term project and the tie may weaken after the project has ended; a couple may break up and be no longer connected in the social network. 
It is therefore more realistic to describe the influence probabilities and social network topology as non-stationary. 
In addition, it is often hard to determine an accurate stochastic distribution assumption for the variance of influence probabilities, since no assumption may exist for human behavior.

To handle unknown underlying distributions in online optimization, multi-armed bandit optimization has been applied in related scenarios. 
The multi-armed bandit problem \cite{auer2002finite} 
is a problem in which an agent has multiple arms to choose from, and needs to decide a policy to select an arm at each time. When chosen, an arm provides a random reward from an unknown distribution specific to the arm, 
and the agent utilizes the outcome to update his strategy. The objective is to maximize the overall reward in the whole time span through selecting a sequence of arms, thus minimize regret, which is the gap between offline optimal overall reward and the actual overall reward the agent has obtained. The design of multi-armed bandit algorithms mainly focuses on how to handle the trade-off between exploration and exploitation \cite{bubeck2012regret}, {\em i.e.}, to try the arm that has not been attempted before ({\em exploration}) or the arm that has brought high reward so far ({\em exploitation}). The basic version of the multi-armed bandit problem considers picking only one arm each time, but there are many extensions, {\em e.g.}, combinatorial bandits \cite{chen2013combinatorial} where multiple arms are chosen at each time. 
Multi-armed bandit optimization has been applied in a wide range of problem domains, including recommendation systems \cite{qin2014contextual}, online wireless channel allocation \cite{gai2012online}, and opportunistic spectrum access \cite{tekin2011online}. 

Multi-armed bandit optimization has also been applied to solve the influence maximization problem with unknown influence probabilities \cite{chen2013combinatorial}\cite{lei2015online}\cite{vaswani2015influence}. 
The existing algorithms have been relying on assumptions of the rewards to guarantee nice theoretical bounds on regret. For example, UCB \cite{bubeck2012best} assumes that the reward distributions are stationary and obtains a regret bound of $O(T)$ under the adversary settings, {\em i.e.}, there is a rival assigning the rewards against the agent. There exist some studies dealing with non-stationary bandits \cite{garivier2011upper}, but none can be readily applied to the influence maximization problem. Detailed discussions of the existing literature in these aspects are given in Sec.~\ref{sec2}.
 
This paper designs an online randomized algorithm, referred to as {\em RSB}, based on multi-armed bandit optimization, to maximize influence propagation in a dynamic non-stationary social network with unknown and non-stationary influence probabilities between pairs of users. 
Our algorithm design does not assume knowledge of the social graph and the influence probability distributions, nor requires any initialization stage. 
Regardless of the concrete influence probabilities or the topology of the social network, an $O(\sqrt{TN\ln N})$ regret bound is rigorously proven where $T$ is the number of time stages in the entire system span and $N$ is the number of nodes. To the best of our knowledge, this is the first influence maximization algorithm dealing with both unknown and non-stationary influence probabilities. We evaluate practical effectiveness of the algorithm using both synthetic and real-world datasets, which demonstrates that our algorithm outperforms previous stationary multi-armed bandit algorithms under non-stationary conditions. 



The rest of the paper is organized as follows. We discuss related work in Sec.~\ref{sec2} and present the problem model in Sec.~\ref{sec3}. In Sec.~\ref{sec4} and Sec.~\ref{sec5}, we present the detailed online algorithm and provide theoretical analysis of its regret bound. Simulation results are presented in Sec.~\ref{sec6}. We conclude the paper in Sec.~\ref{sec7}.
\section{Related Work}\label{sec2}

\subsection{Influence Maximization with Bandit Optimization}
\label{bandit_influ_max}

Recently, multi-armed bandit optimization has been applied to solve influence maximization problem with incomplete information of the social network. 
In particular, combinatorial bandits are highly correlated to the influence maximization problem, where the decision-making agent needs to select multiple arms in each time stage. Chen \textit{et al.}~\cite{chen2013combinatorial} define a general bandit framework to deal with both linear and non-linear reward functions. Applying the upper confidence bound approach, they achieve a regret with a mild logarithmic dependence on the total number of time stages. One application of their framework is social influence maximization with unknown influence probabilities. However, the reward of each arm must be an i.i.d random process, {\em i.e.}, the reward distribution is stationary over time. Lei \textit{et al.}~\cite{lei2015online} present an online influence maximization framework utilizing exploration-exploitation for seed selection and strategy updating. They assume that all connections in the social network are known and only evaluate the performance of their algorithm experimentally without any theoretical analysis. 

Targeting news delivery, Massouli{\'e} \textit{et al.}~\cite{massoulie2015greedy} propose a greedy Bayesian approach in bandit optimization to identify all posteriori interested users for delivering fresh news of unknown topics, spamming least uninterested users at the same time. 
Their regret bound is logarithmical with the number of users. In each time stage, the news is pushed to only one user, and their approach cannot be readily extended to combinatorial scenarios. 
Yue \textit{et al.}~\cite{yue2011linear} model personalized news recommendation using linear submodular bandits under stationary probability distributions. 
Lin \textit{et al.}~\cite{lin2015stochastic} develop a more general bandit optimization algorithm for a class of problems using greedy methods, guaranteeing a good approximation ratio. However, they still only consider stochastic rewards with stationary distributions.
 
\subsection{Multi-Armed Bandit with Non-Stationary Rewards}
\label{bandit_nonstationary}
The simplest idea to tackle non-stationary rewards is to decrease the weights of earlier feedback in next-step decision making \cite{garivier2008upper}. The problem it may lead to is that without sufficient feedback information, it is hard to achieve a good accuracy of reward estimation. Some algorithm designs assume abrupt changes of the distributions occurring at arbitrary intervals \cite{yu2009piecewise}, and allow the agent to query a set of arms not picked before and obtain outcomes as if these arms were played. 
This assumption is reasonable in a stock market, {\em i.e.}, people can acquire information of stocks they have never purchased by following bearish or bullish trends, but not for influence propagation, where there is no channel to obtain outcomes of untried arms. Besbes \textit{et al.}~\cite{besbes2014stochastic} assume that the total variation of the 
rewards is given and design a randomized algorithm based on Exp3, which assigns exponential weights to arms for exploration and exploitation in adversary bandit \cite{auer2003nonstochastic}. Only one arm is selected in each time stage, while we focus on the case of combinatorial bandits. It is non-trivial to extend the algorithm to combinatorial scenarios. 

Gai \textit{et al.}~\cite{gai2011combinatorial} study non-stationary bandit optimization under the assumption that the state of a selected arm evolves as an irreducible finite-state Markov process with unknown transition matrix, while the distributions of other arms stay unchanged (rested arms). Their work is applicable to many graph theory problems, {\em e.g.}, channel allocation in cognitive radio networks. However, assuming rested arms is not realistic in influence propagation in social networks. 
The same authors further investigate restless bandits with Markov rewards \cite{gai2012online}, where the states of an arm evolve dynamically over time no matter whether it has been played. The algorithm utilizes regenerative property of a Markov chain and achieves a regret near logarithmic on the total number of time stages. Both studies rely on an initialization stage, in which each arm is tried for at least once. This is impractical for influence propagation ({\em e.g.}, market campaign) in a large-scale network, as the cost of trying all nodes is unaffordable. 
Granmo \textit{et al.}~\cite{granmo2010solving} use Kalman filter to update estimation of the reward distribution, and evaluate their results by simulation without theoretical analysis. Kalman filter is only applicable to linear dynamic system and the states inherently form a Markov chain. It is not realistic to make the Markov chain assumption in influence propagation, since human behavior does not simply depend on one's latest status.
\section{Problem Model}\label{sec3}

We model the social network as an influence graph $G=(\mathcal{N}, \mathcal{E})$. $\mathcal{N}=\{1,2,\dots,N\}$ is the set of users (nodes), where $N$ is the total number of nodes. $\mathcal{E}$ is the set of social connections among the nodes. An unknown influence probability $p_{n,m}^t$ is associated with each edge $(n,m)\in\mathcal{E}$, which is time varying following an unknown, non-stationary distribution: after user $n$ is activated ({\em e.g.}, obtained information of a product), he may activate his neighbor $m$ ({\em e.g.}, share information of the product) with different probabilities at different time stages $t$. 
In this way, each edge $(n,m)$ is associated with a non-stationary Bernoulli distribution: in $t$, user $n$ may activate his neighbor $m$ with probability $p_{n,m}^t$, or not with probability $1-p_{n,m}^t$. We do not assume any cascade model of the information propagation system ({\em e.g.}, independent cascade model or linear threshold model), and our algorithm works with various cascade models as long as the information spread brought by an activate node can be modeled as a random variable.

Let $T$ be the total number of time stages that the system spans. In each time stage, a set of $K$ seeds are selected as information sources ({\em e.g.}, the seed users that an advertiser directly promotes the product to, whose number is decided by the promotion budget), from which the information spreads to other nodes in the network. The seed set is repeatedly selected over different time stages. 
For example, a company may carry out a promotion campaign for a series of time stages, {\em e.g.}, a number of consecutive days. After the promotion in each time stage via a potentially different set of seeds, the company collects statistics on the number of purchases of their promoted product(s) and utilizes this feedback to update its seed selection strategies in later time stages. The goal is to maximize the expected overall influence spread in the whole time span $1,2,\ldots, T$, {\em i.e.}, the expected total number of activated nodes. 
Let $\mathcal{M}$ be the collection of all subsets of $\mathcal{N}$. In our bandit optimization framework, we define $a \vert S$, meaning node $a$ under a given set $S \in \mathcal{M}$, as an {\em arm}. The expected {\em reward} of selecting an arm $a \vert S$ is the expected marginal gain by adding $a$ into the existing seed set $S$, {\em i.e.}, the expected additional number of activated users after we add $a$ into $S$. Let $f_t(S)$ 
be an influence spread function in time stage $t$, indicating the total number of activated nodes in $t$ based on seed set $S$. 
The value of $f_t(S)$ is a random variable. The expectation $\mathbb{E} [f_t(S)]$ is non-negative, monotone and submodular, as proven in \cite{chen2013information}. The submodularity of the spread function is useful such that we can utilize the benchmark based on greedy optimal value. The expected reward of selecting an arm $a \vert S$ in $t$ is hence $\mathbb{E}[f_t(S \cup \{a\})]-\mathbb{E}[f_t(S)]$. Note that the expectation $\mathbb{E}[\cdot]$ is taken over both randomized rewards and randomized policies, where a {\em policy} refers to the agent's strategy for seed selection, which is random given the random nature of our algorithm.\footnote{Although \cite{chen2013information} does not consider randomized policies, the submodularity of $\mathbb{E} [f_t(S)]$ still holds following results in \cite{chen2013information}, as expectation over policies is a linear combination of submodular functions.}

In each time stage $t$, starting from an empty set $S_t=\emptyset$, we obtain a seed set of size $K$ by adding nodes to $S_t$ one by one in some order. Let $S_t=(a_t^1,\dots,a_t^K)$ be the completed seed set, in which the $k\textsuperscript{th}$ element is the $k\textsuperscript{th}$ seed selected in this time stage. 
Let $S_t^{(1:k-1)}$ represent the selected seed set with elements $1,\dots,k-1$, and $a_t^k \vert S_t^{(1:k-1)}$ mean that node $a_t^k$ is selected as the $k\textsuperscript{th}$ seed in $t$ given previous choices in $S_t^{(1:k-1)}$. Let $\bar{r}^k_t(a_t^k \vert S_t^{(1:k-1)})=\mathbb{E}[f_t(S_t^{(1:k-1)} \cup \{a\})]-\mathbb{E}[f_t(S_t^{(1:k-1)})]$ denote the expected marginal gain of choosing $a_t^k$ as the $k\textsuperscript{th}$ seed in $t$. The expected total reward in time stage $t$ is $\bar{r}_t(S_t)=\sum_{k=1}^K \bar{r}^k_t(a_t^k \vert S_t^{(1:k-1)})=\mathbb{E}[f_t(S_t)]$. 

In this model, maximizing the expected total number of activated nodes in $1, \ldots, T$ is equivalent to maximizing the expected overall reward in the entire span, $\sum_{t=1}^T\bar{r}_t(S_t)=\sum_{t=1}^T\mathbb{E}[f_t(S_t)]$. It is further equivalent to minimizing the regret, the gap between the expected overall reward that the agent can obtain by running our online algorithm and the offline optimal expected overall reward computed using full knowledge of the system. In our algorithm design, we aim to minimize the weak regret, {\em i.e.}, the gap between the expected overall reward achieved by our algorithm and the offline expected overall reward achieved by using the same best seed set $S^*$ in all time stages, namely $S^* \in  \underset{S \in \mathcal{M}}{\operatorname*{arg\,max}} \sum_{t=1}^T \mathbb{E}[f_t(S)]$, computed based on full knowledge of the entire system. Such a weak regret is the difference between the expected overall reward obtained by our algorithm and that achieved by the best single action, {\em i.e.}, sticking with one seed set in all time stages. Weak regret is commonly used in the literature on analysing non-stationary bandit algorithms \cite{gai2011combinatorial}\cite{gai2012online}\cite{tekin2011online}\cite{liu2013learning}, and 
the key ingredient is to form accurate estimates on the average condition for each arm \cite{liu2015online}, so as to find the arm performing best in a long term. 
In particular, we analyze a greedy weak regret, with detailed definition given in Definition \ref{def3} in Sec.~\ref{sec5}, that compares the expected overall reward produced by our algorithm with the lower bound of an approximate offline overall reward achieved by a single best seed set derived by a greedy approach. Greedy weak regret is a concept narrowed down from weak regret, when the best single action is decided by a greedy algorithm. We apply this notion so as to compare with the lower bound of the greedy optimal value.
\section{RSB: Randomized Multi-armed Bandit Algorithm for Non-Stationary Social Networks}\label{sec4}

\begin{table}
\renewcommand{\arraystretch}{1.3}
\caption{Notation}\label{table1}
\begin{center}
\begin{tabular}{| c | p{6cm} |}
\hline
$N$ & $\#$ of nodes\\ \hline
$\mathcal{N}$ & the set of nodes\\ \hline
$\mathcal{M}$ & the collection of subsets of $\mathcal{N}$\\ \hline
$T$ & the total number of time stages\\ \hline
$C$ & input parameter to Alg.~\ref{algo2}\\ \hline
$K$ & the size of seed set\\ \hline
$\gamma$ & input parameter to Alg.~\ref{algo2}\\ \hline
$S_t^{(1:k-1)}$ & the set containing the first $k-1$ selected seeds in $t$\\ \hline
$a \vert S_t^{(1:k-1)}$ & an arm in $t$, selecting node $a$ given $S_t^{(1:k-1)}$\\ \hline
$f_t(S)$ & the influence spread of seed set $S$ in $t$\\ \hline
$r_t^k(a \vert S_t^{(1:k-1)})$ & reward of choosing $a$ as $k\textsuperscript{th}$ seed based on $S_t^{(1:k-1)}$ in $t$\\ \hline
$\bar{r}_t^k(a \vert S_t^{(1:k-1)})$ & expected reward of choosing $a$ as $k\textsuperscript{th}$ seed based on $S_t^{(1:k-1)}$ in $t$\\ \hline
$Reg_G(T)$ & greedy weak regret in the whole system span\\ \hline
$Reg^k(t)$ & position weak regret for the $k\textsuperscript{th}$ seed in $t$\\ \hline
$a_t^k$ & selected node as $k\textsuperscript{th}$ seed in $t$\\ \hline
$\tilde{a}^k$ & optimal node as $k\textsuperscript{th}$ seed in all time stages\\ \hline
$\mathbb{E}[\cdot]$ & expectation taken over both random policies and random rewards\\ \hline
$OPT$ & the offline maximal value of the expected overall reward\\ \hline
\end{tabular}
\end{center}
\end{table}

\noindent \textbf{Main Idea.} We next design an online multi-armed bandit algorithm to minimize the greedy weak regret. In each time stage, we select the best seed set by sequentially selecting the next best node given previous seed decisions. Given the set of already selected seeds, we associate weights with candidate arms, and deal with the varying environment (time-varying underlying distributions of influence probabilities) by adjusting the weights of arms based on rewards received due to previous seed selection (the {\em exploitation} component of our algorithm). Besides, we also include a constant $\frac{\gamma}{N}$ in the weight of each arm, where $\gamma \in (0,1]$ is a gaugeable value, in order to enable {\em exploration} of arms never tried before. Different from deterministic stationary bandit algorithms, our algorithm is randomized in arm selection according to the weights, and hence even if the environment changes abruptly, the algorithm still has a chance to switch to the new best arm.

 
\noindent \textbf{Algorithm Steps.} Our multi-armed bandit algorithm for selecting the best seed set in each time stage $t$ is given in Alg.~\ref{algo2}. Here $w_t^{n \vert S_t^{(1:k-1)}}$ is the weight for selecting node $n$ as the $k\textsuperscript{th}$ seed in time stage $t$, while the set of already selected seeds in $t$ is $S_t^{(1:k-1)}$. $v_t^{n \vert S_t^{(1:k-1)}}$ is an auxiliary quantity to compute the weights, updated based on the past reward information of arm $n \vert S_t^{(1:k-1)}$, as an exploration measure. 
$q_t^{n \vert S_t^{(1:k-1)}}$ is the probability of playing arm $n \vert S_t^{(1:k-1)}$ in $t$,  derived from the weights of the arms. 
$r_t^k(a \vert S_t^{(1:k-1)})$ denotes the realization of the reward (actual marginal influence spread) by choosing node $a$ as the $k\textsuperscript{th}$ seed in $t$. $C$ is an input parameter to the algorithm, which satisfies $C\geq \frac{\gamma r_t^k(n \vert S)}{N q_t^k(n \vert S)}, \forall n \in \mathcal{N},\ S \in \mathcal{M}$.


In Alg.~\ref{algo2}, the $K$ seeds are selected sequentially (line \ref{line2}). The weights $\mathbf{w}$ associated with the nodes should be equal at the beginning of each time stage, and adjusted based on updated $\mathbf{v}$, each time after the seed set has been updated (lines \ref{line3}-\ref{line5}). The computation of $w_t^{n \vert S_t^{(1:k-1)}}$ aims to balance exploitation and exploration: the first term is calculated based on past reward information ({\em exploitation}) and the second constant term is assigned for each arm no matter how many times it has been tried ({\em exploration}). Next, the probability for adding an additional node into the already selected set of seeds is decided by normalizing its weight over the weights of all the remaining nodes not in the existing seed set (lines \ref{line6}-\ref{line8}). An arm is randomly selected according to the probability distribution and a reward $a \vert S_t^{(1:k-1)}$ is observed (lines \ref{line9}-\ref{line11}), {\em e.g.}, the additional number of product purchases received by promoting to node $a$ is collected. We then update $v_t^{n \vert S_t^{(1:k-1)}}$ by multiplying an exponential factor (line \ref{line16}), decided by $\hat{r}_t^k(a \vert S_t^{(1:k-1)})$, which can be understood as an unbiased estimation of the reward of the arm. Computing $\hat{r}_t^k(a \vert S_t^{(1:k-1)})$ by dividing the actual reward by the probability of selecting the arm (line \ref{line12}) compensates the reward of actions with less probability to be chosen and guarantees that the expectation of the estimated reward and the actual reward are equal, when the expectation is taken over both randomized policies and randomized rewards. This equality helps us to derive the expected reward of RSB in the proof.
 The updated weights will be used in selecting future seeds in this time stage.


We will evaluate the impact of the input parameter $\gamma$ under practical settings in simulations. The input parameter $C$ is related to the largest spread brought by a seed, which is unknown before running the algorithm. In fact, requiring $C \geq \frac{\gamma r_t^k(n \vert S)}{N q_t^k(n \vert S)}$ is only needed for regret analysis. We can set the value of $C$ empirically when running the algorithm in practice, and will evaluate the performance of the algorithm under an empirical value of $C$ in simulations, which does not necessarily satisfies the above condition.


\begin{algorithm}[!t]
\caption{RSB: Randomized Sequential Multi-armed Bandit Algorithm for Non-Stationary Networks}
\label{algo2}
\begin{algorithmic}[1]
   	\Require $\mathcal{N}$, $K$, $C$, $\gamma$
   	\Ensure the seed set $S_t^{(1:K)}$ for each time stage $t$
	\State set $v_{1}^{n \vert S_{1}^{(1:k-1)}}=1$, $\forall n \in \mathcal{N}$, $k=1,\dots,K$ 
    	\For{$t=1,2,\dots,T$}
    		\For{$k=1,2,\dots,K$} \label{line2}
    			\For {each node $n \in \mathcal{N}$} \label{line3}
    				\State set $w_t^{n \vert S_t^{(1:k-1)}}=(1-\gamma)\frac{v_t^{n \vert S_t^{(1:k-1)}}}{\sum \limits_{n' \in \mathcal{N}} v_t^{n' \vert S_t^{(1:k-1)}}}+\frac{\gamma}{N}$ \label{linew}
    			\EndFor \label{line5}
    			\For {each node $n \in \mathcal{N} \backslash S_t^{(1:k-1)}$} \label{line6}
    				\State $q_t^{n \vert S_t^{(1:k-1)}}=\frac{w_t^{n \vert S_t^{(1:k-1)}}}{\sum \limits_{n' \in \mathcal{N}\backslash S_t^{(1:k-1)}} w_t^{n' \vert S_t^{(1:k-1)}}}$
    			\EndFor \label{line8}
    			\State draw an arm $a \vert S_t^{(1:k-1)}$ according to the distribution $\{q_t^{n \vert S_t^{(1:k-1)}}\}_{n \in \mathcal{N} \backslash S_t^{(1:k-1)}}$ \label{line9}
    			\State receive a reward $r_t^k(a \vert S_t^{(1:k-1)})$
    			\State set $S_t^{(1:k)}=S_t^{(1:k-1)} \cup \{a\}$ \label{line11}
    			\State 
				set $\hat{r}_t^k(a \vert S_t^{(1:k-1)})=\frac{r_t^k(a \vert S_t^{(1:k-1)})}{q_t^{a \vert S_t^{(1:k-1)}}}$ \label{line12}
    			\State for all $n \in \mathcal{N}\backslash \{a\}$, set $\hat{r}_t^k(n \vert S_t^{(1:k-1)})=0$
    			\For{each arm $n\vert S_t^{(1:k-1)},\ \forall n \in \mathcal{N}$}
    				\State update $v_{t+1}^{n \vert S_{t+1}^{(1:k-1)}}=v_t^{n \vert S_t^{(1:k-1)}}\exp{\{\frac{\gamma \hat{r}_t^k(n \vert S_t^{(1:k-1)})}{NC}\}}$ \label{line16}
    			\EndFor 
    		\EndFor
	\EndFor
\end{algorithmic}
\end{algorithm}

We note that our algorithm does not rely on any knowledge of the underlying social network topology and the influence probabilities, but only utilizes the outcomes that are decided by them. In addition, although the entire space of arms, $a \vert S, \forall a \in \mathcal{N},\ S \in \mathcal{M}$, is exponential, the number of arms that need to be dealt with in each time stage in Alg.~\ref{algo2} (weights and probabilities computed and used in seed selection) is still polynomial, as given in the following theorem. 

\begin{theorem}
The time complexity of Alg.~\ref{algo2}, executed in each time stage $t$, is $O(KN)$. 
\end{theorem}
\begin{IEEEproof}
In each time stage $t$, we select $K$ seeds. When selecting the $k\textsuperscript{th}$ seed based on already selected seeds in $S_t^{(1:k-1)}$, we compute/update weights, and compute selection probabilities for at most $N$ arms corresponding to $N$ nodes in the network. Therefore, the time complexity is $O(KN)$.
\end{IEEEproof}

\section{Regret Analysis}\label{sec5}

We next analyze an upper bound of the greedy weak regret achieved by Alg.~\ref{algo2}. 
Let $OPT$ denote the offline maximal value of the expected overall reward $\sum_{t=1}^T\bar{r}_t(S)=\sum_{t=1}^T \mathbb{E} [f_t(S)]$ over all $S \in \mathcal{M}$, 
computed based on complete knowledge of the influence probability distributions and the social graph topologies in $1, \ldots, T$. 
Let $S^*$ be the offline optimal seed set, {\em i.e.}, the single best seed set that maximizes $\sum_{t=1}^T\bar{r}_t(S)$. 
 
\subsection{Reduction from Greedy Weak Regret to Position Weak Regret}

We define a {\em position} optimal reward $OPT^k$ as the sum of the expected marginal gains achieved by using the best $k\textsuperscript{th}$ seed in all time stages. The best $k\textsuperscript{th}$ seed maximizes $\sum_{t=1}^T\bar{r}_t^k(a \vert S_t^{(1:k-1)})$ based on full knowledge of the system, 
given the first $k-1$ seeds in $S_t^{(1:k-1)}$ in each $t$ derived using RSB. 
The idea is to reduce the original problem of finding the best solution of the full set to a parallel bandit setting, finding the best $k\textsuperscript{th}$ element under the condition determined by our algorithm. 
Let $\tilde{a}^k$ denote this optimal $k\textsuperscript{th}$ seed, {\em i.e.}, $\tilde{a}^k\in \underset{a \in \mathcal{N}}{\operatorname*{arg\,max}} \sum_{t=1}^{T} \bar{r}_t^k (a \vert S_t^{(1:k-1)})$. 
Such a best $k\textsuperscript{th}$ seed may form different arms, $\tilde{a}^k \vert S_t^{(1:k-1)}$, under different seed sets $S_t^{(1:k-1)}$ in different time stages. We have $OPT^k=\max \limits_{a \in \mathcal{N}} \sum_{t=1}^T \bar{r}_t^k(a \vert S_t^{(1:k-1)})=\sum_{t=1}^T \bar{r}_t^k (\tilde{a}^k \vert S_t^{(1:k-1)})$.
\begin{definition}\label{position_regret}
The position weak regret for the $k\textsuperscript{th}$ seed is
\begin{align*}
Reg^k(T)=\sum_{t=1}^T \bar{r}_t^k (\tilde{a}^k \vert S_t^{(1:k-1)})-\sum_{t=1}^T \bar{r}_t^k (a^k_t \vert S_t^{(1:k-1)})
\end{align*}
where $\tilde{a}^k\in \underset{a \in \mathcal{N}}{\operatorname*{arg\,max}} \sum_{t=1}^{T} \bar{r}_t^k (a \vert S_t^{(1:k-1)})$ and $a^k_t \vert S_t^{(1:k-1)}$ is the arm selected by Alg.~\ref{algo2} in time stage $t$. The conditional set $S_t^{(1:k-1)}$ is also decided by Alg.~\ref{algo2}.
\end{definition}




The following theorem states the relationship between position weak regret and $OPT$, which will be used to bound the greedy weak regret in Theorem \ref{th3}. Its proof can be found in Appendix \ref{sec:th2}.

\begin{theorem} \label{th2}
For any position $k=1,2,\ldots,K$, we have
\begin{align*}
& \sum_{t=1}^T \Big( \bar{r}_t(S_t^{(1:k)})-\bar{r}_t(S_t^{(1:k-1)}) \Big)\\
& \geq \frac{1}{K}\Big( OPT-\sum_{t=1}^T \bar{r}_t(S_t^{(1:k-1)})\Big)-Reg^k(T).
\end{align*}
\end{theorem}

Let $F(S)=\sum_{t=1}^T \mathbb{E}[f_t(S)]$, $\forall S \in \mathcal{M}$, which denotes the expected overall influence spread over the whole system span. $F(S)$ is a submodular function since it is the summation of submodular functions $\mathbb{E}[f_t(S)],\forall t=1,\ldots, T$. Then we can design a greedy approach to compute a $S$ that approximately maximizes the expected overall reward $\sum_{t=1}^T\bar{r}_t(S)=\sum_{t=1}^T \mathbb{E} [f_t(S)]$ based on full knowledge of the system: after deciding $S^{(1:k-1)}$, we select a local optimal node as the $k\textsuperscript{th}$ seed, that maximizes the expected marginal influence spread, {\em i.e.}, node $u$ such that $u \in \underset{v \in \mathcal{N} \backslash S^{(1:k-1)}}{\operatorname*{arg\,max}} \{F(S^{(1:k-1)} \cup \{v\})-F(S^{(1:k-1)})\}$. 
We can easily prove that the approximate offline solution computed this way achieves an approximation ratio of ${1-\frac{1}{e}}$, {\em i.e.}, the expected overall reward it achieves is at least $(1-\frac{1}{e})OPT$, following Theorem 3.5 in \cite{chen2013information}, based on submodularity of the spread function and local optimality when selecting each seed. 
The reason that we compute this approximate offline solution using the greedy approach (which runs in polynomial time) is that computing $S^*$ has been shown to be an NP hard problem \cite{kempe2003maximizing}.


Using the approximate offline overall reward computed as above, we define a greedy weak regret as follows, which we use to evaluate the performance of our algorithm RSB.

\begin{definition} \label{def3}
The greedy weak regret is defined as the gap between the lower bound of the approximate offline overall reward derived by the greedy approach 
and the expected overall reward produced by RSB in Alg.~\ref{algo2}, {\em i.e.},
\begin{align*}
Reg_G(T)=(1-\frac{1}{e}) OPT -\sum_{t=1}^T \bar{r}_t(S_t^{(1:K)}).
\end{align*}
\end{definition}

The following theorem shows that the overall position weak regret provides an upper bound of the greedy weak regret. The proof can be found in Appendix \ref{sec:th3}.
\begin{theorem} \label{th3}
The greedy weak regret is upper bounded by the sum of position weak regrets over all positions $k=1, 2, \ldots, K$, {\em i.e.},
\begin{align*}
Reg_G(T) \leq \sum_{k=1}^K Reg^k(T). 
\end{align*}
\end{theorem}

Based on Theorem \ref{th3}, we seek to bound the position weak regret for each $k$, in order to derive an upper bound of $Reg_G(T)$.


\subsection{Bounding Greedy Weak Regret}

According to Definition \ref{position_regret}, the position weak regret for the $k\textsuperscript{th}$ seed is 
\begin{align*}
Reg^k(T)
& =\sum_{t=1}^{T} r_t^k (\tilde{a}^k \vert S_t^{(1:k-1)})-\sum_{t=1}^{T} r_t^k(a^k_t \vert S_t^{(1:k-1)})\\
&=\max \limits_{n \in \mathcal{N}} \sum_{t=1}^{T} r_t^k (n \vert S_t^{(1:k-1)})-\sum_{t=1}^{T}  r_t^k(a^k_t \vert S_t^{(1:k-1)}).
\end{align*}

%
%
%
%
%
%

Let $D$ be the upper bound of the realization of reward, {\em i.e.}, $r_t^k(n \vert S) \leq D,\  \forall n \in \mathcal{N},\ S \in \mathcal{M}$. The following theorem states an upper bound of the position weak regret for each $k$. 
In particular, if $\gamma$ is set to a special value, it can minimize the regret bound. 
The proof can be found in Appendix \ref{sec:th4}.
\begin{theorem}\label{th4}
Let $R^k_{\max}=\max \limits_{n \in \mathcal{N}} \sum_{t=1}^{T} \bar{r}_t^k (n \vert S_t^{(1:k-1)})$ be the expected overall reward achieved by selecting the best $k\textsuperscript{th}$ arm given $S_t^{(1:k-1)}, \forall t=1, \ldots, T$, derived by Alg.~\ref{algo2}. Let $R^k_{RSB}=\sum_{t=1}^{T}  \mathbb{E}[r_t^k(a^k_t \vert S_t^{(1:k-1)})]$ denote the expected overall marginal gain obtained by adding the $k\textsuperscript{th}$ seeds into the given $S_t^{(1:k-1)}, \forall t=1, \ldots, T$. 
For any parameter $\gamma \in (0,1]$, we have
\begin{align*}
& Reg^k(T) = R^k_{\max}-R^k_{RSB}\\
& \leq (1+(e-2)\frac{D}{C}) \gamma  R^k_{\max} +\frac{NC \ln N}{\gamma}.%
\end{align*}
If we set $\gamma=\min\{1,\sqrt{\frac{NC\ln N}{(1+(e-2)\frac{D}{C})g}}\}$ where constant $g \geq  R^k_{\max},\ \forall k=1, \ldots, K$, we have the following minimum upper bound
\begin{align*}
\sum_{k=1}^K Reg^k(T) \leq 2K \sqrt{1+(e-2)\frac{D}{C}} \sqrt{gCN \ln N}.
\end{align*}
\end{theorem}

\newtheorem{corollary}{Corollary}
\begin{corollary} \label{coro1}
The greedy weak regret achieved by Alg.~\ref{algo2} is upper bounded as follows:
\begin{align*}
Reg_G(T) \leq 2K \sqrt{1+(e-2)\frac{D}{C}} \sqrt{DCTN \ln N},
\end{align*}
{\em i.e.}, the upper bound of the greedy weak regret of Alg.~\ref{algo2} is $O(\sqrt{TN\ln N})$.
\end{corollary}
It shows that our greedy weak regret is sublinear with both $N$ and $T$. The proof can be found in Appendix \ref{sec:coro1}.
\section{Performance Evaluation}\label{sec6}

\subsection{Data Sets}

\subsubsection{Tencent Weibo Traces}
We produce a dynamic social network based on Tencent Weibo\footnote{http://t.qq.com/} traces containing the {\em following} links among $4257$ users for $7$ consecutive days during November $2011$. Each directed {\em following} link $(n,m)$ indicates that user $n$ follows user $m$ \cite{yi2014building}. 
The links among the users vary from one day to the next, giving a dynamic social graph. 
To prolong the trace duration, we further repeat the variation of the social graph on $7$-day cycles to form a $100$-day duration ($T$), which we believe reasonable since human behavior may well follow a weakly periodicity.

\subsubsection{Synthetic Data}
As Weibo traces only provide the dynamic behavior of a specific social network, we also generate a synthetic dynamic social network by combining the model in \cite{zhuang2013influence} with the Erd\H{o}s-R\'enyi model and preferential attachment: we generate an initial graph with $5000$ nodes and connect each pair of node with probability $0.005$ (a directed link); then in each time stage, we select $1000$ edges uniformly and change their heads to other nodes randomly picked with probabilities proportional to their indegrees. In this way, we generate a sequence of social graphs for $T=100$ time stages. Preferential attachment is a representative mechanism to model the topology of a social network, that the more connected a node is, the more likely it is to receive new links.

\subsection{Time-varying Influence Probabilities}
We employ the following three models to generate nonuniform and time-varying influence probabilities in a social graph. 
\begin{itemize}
\item The Weighted Cascade (WC) model \cite{kempe2003maximizing}: the influence probability $p_{n,m}^t$ of edge $(n,m)$ at time $t$ is $\frac{1}{d_m^t}$, where $d_m^t$ is the indegree of node $m$ at $t$. The probabilities are varying due to the changes of links in a dynamic social graph.

\item The Trivalency (TR) model \cite{chen2010scalable}: in each time stage, the influence probability of an edge in the social graph is assigned a value among $\{0.1,\ 0.01,\ 0.001\}$ uniformly randomly, corresponding to three types of social ties - strong, medium and weak. The assigned probability on an edge may change from one time stage to the next.

\item A Fluctuating Reward (FR) model. We design this model such that influence probabilities evolve over time in a similar fashion as a sinusoidal wave (also similar to that used in \cite{besbes2014optimal}): the influence probability of each edge starts from a random value drawn uniformly within $[0, 0.1]$; then in each time stage, it increases or decreases at a constant rate $\frac{0.3}{T}$ until reaching the largest value $0.1$ or the smallest value $0$. 
\end{itemize}

\subsection{Schemes for Comparison}
We compare RSB with a random algorithm and OG-UCB proposed in \cite{lin2015stochastic}. With the random algorithm, the agent always selects a seed uniformly randomly among all candidate nodes. OG-UCB is designed for stationary scenarios, which associates a confidence bound with each arm and chooses the arm with the highest upper confidence bound greedily.

We note that although a number of bandit algorithms have been proposed for influence maximization (as discussed in Sec.~\ref{bandit_influ_max}), most are not directly comparable since they run with the complete knowledge of a social network. We compare with OG-UCB since it is the only existing bandit algorithm without requiring knowledge of the social graph topology. 
In addition, the bandit algorithms designed for non-stationary systems in Sec.~\ref{bandit_nonstationary} either deal with 1 arm or assume Markov rewards, and hence cannot be readily extended for comparison.
 
In computing greedy weak regret, we also compute the approximate offline optimal overall reward by the greedy offline algorithm discussed before Definition \ref{def3} in Sec.~\ref{sec5}. 

\subsection{Evaluation Results}

To show greedy weak regret values in a unified range in our figures, we plot the ratio between greedy weak regret and the approximate offline optimal overall reward, {\em i.e.}, $\frac{\mbox{approx. offline opt. overall reward}-\mbox{overall reward by RSB}}{\mbox{approx. offline opt. overall reward}}$. Especially, a data point at a specific $T$ represents the above ratio computed using overall rewards in $[1,T]$. We set $K=5$, $\gamma=0.2$ (default), $D=120$ and $C=1$ in our experiments. 

\begin{figure*}[!htp]
\begin {center}
\begin{minipage}[t]{0.32\linewidth}
\centering
\includegraphics[width=\linewidth]{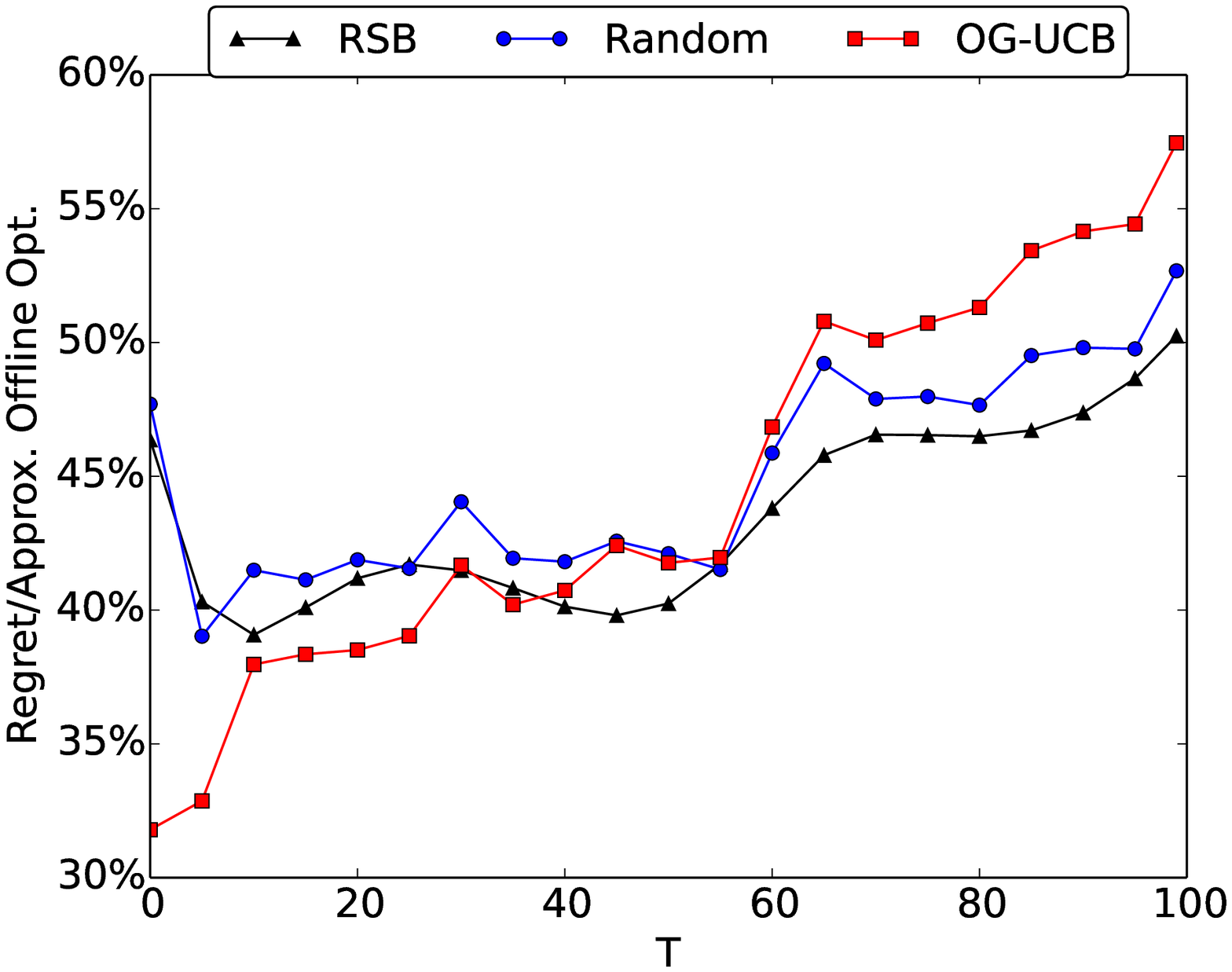}
\caption{Synthetic data and WC model.}
\label{fig1}
\end{minipage}
\hfill
\begin{minipage}[t]{0.32\linewidth}
\centering
\includegraphics[width=\linewidth]{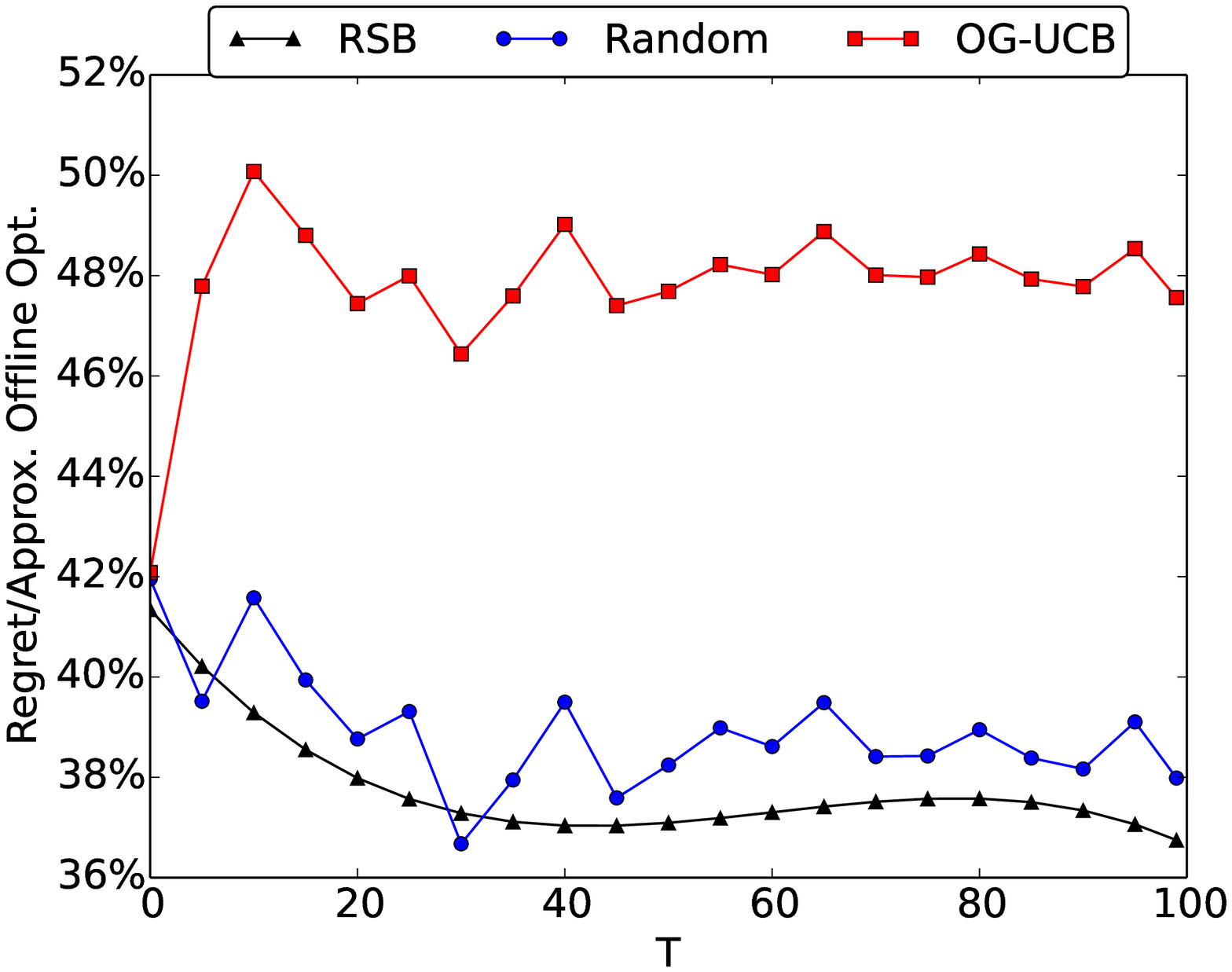}
\caption{Synthetic data and TR model.}
\label{fig2}
\end{minipage}
\hfill
\begin{minipage}[t]{0.32\linewidth}
\centering
\includegraphics[width=\linewidth]{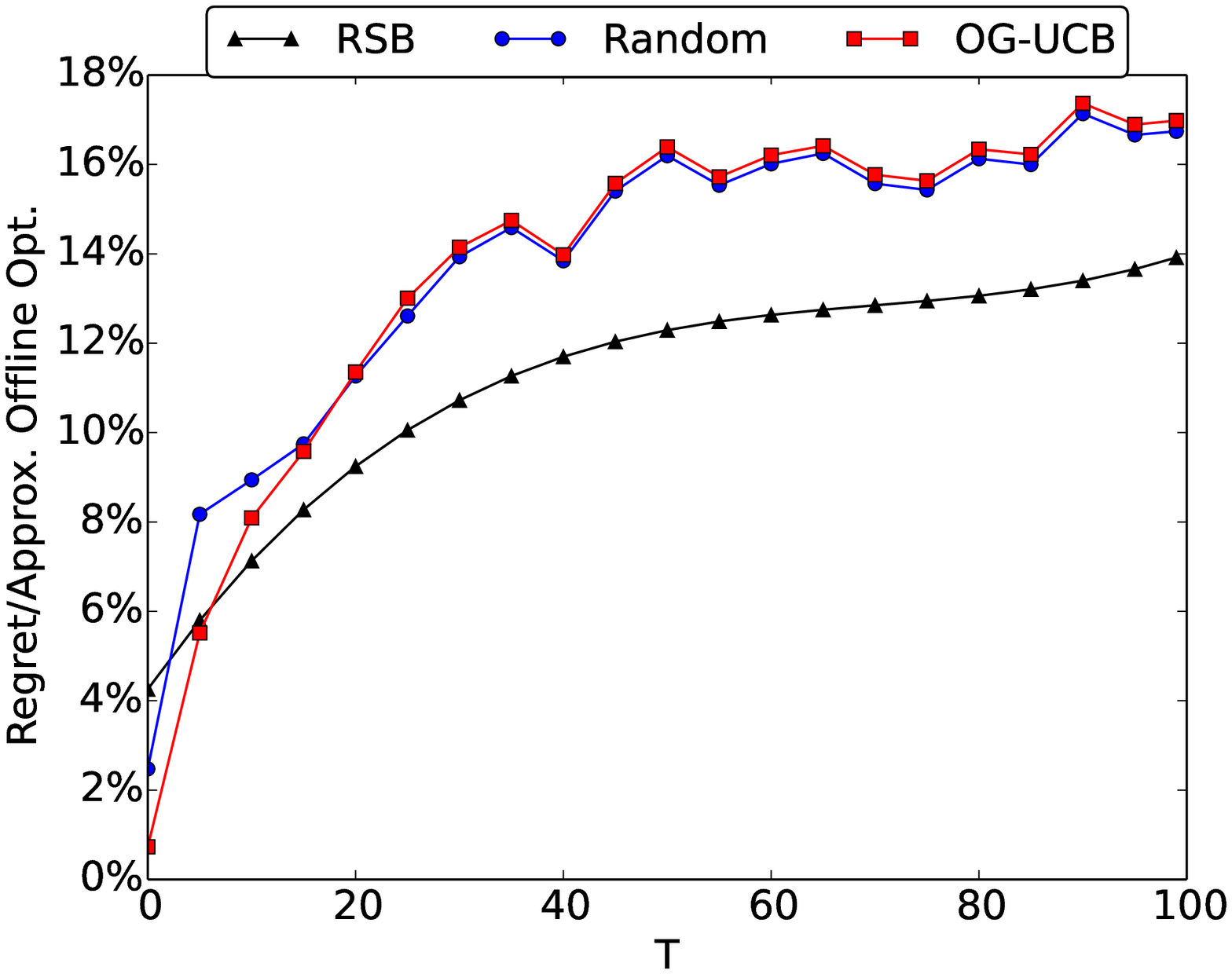}
\caption{Synthetic data and FR model.}
\label{fig3}
\end{minipage}
\end{center}
\end{figure*}

\begin{figure*}[!htp]
\begin{center}
\begin{minipage}[t]{0.32\linewidth}
\centering
\includegraphics[width=\linewidth]{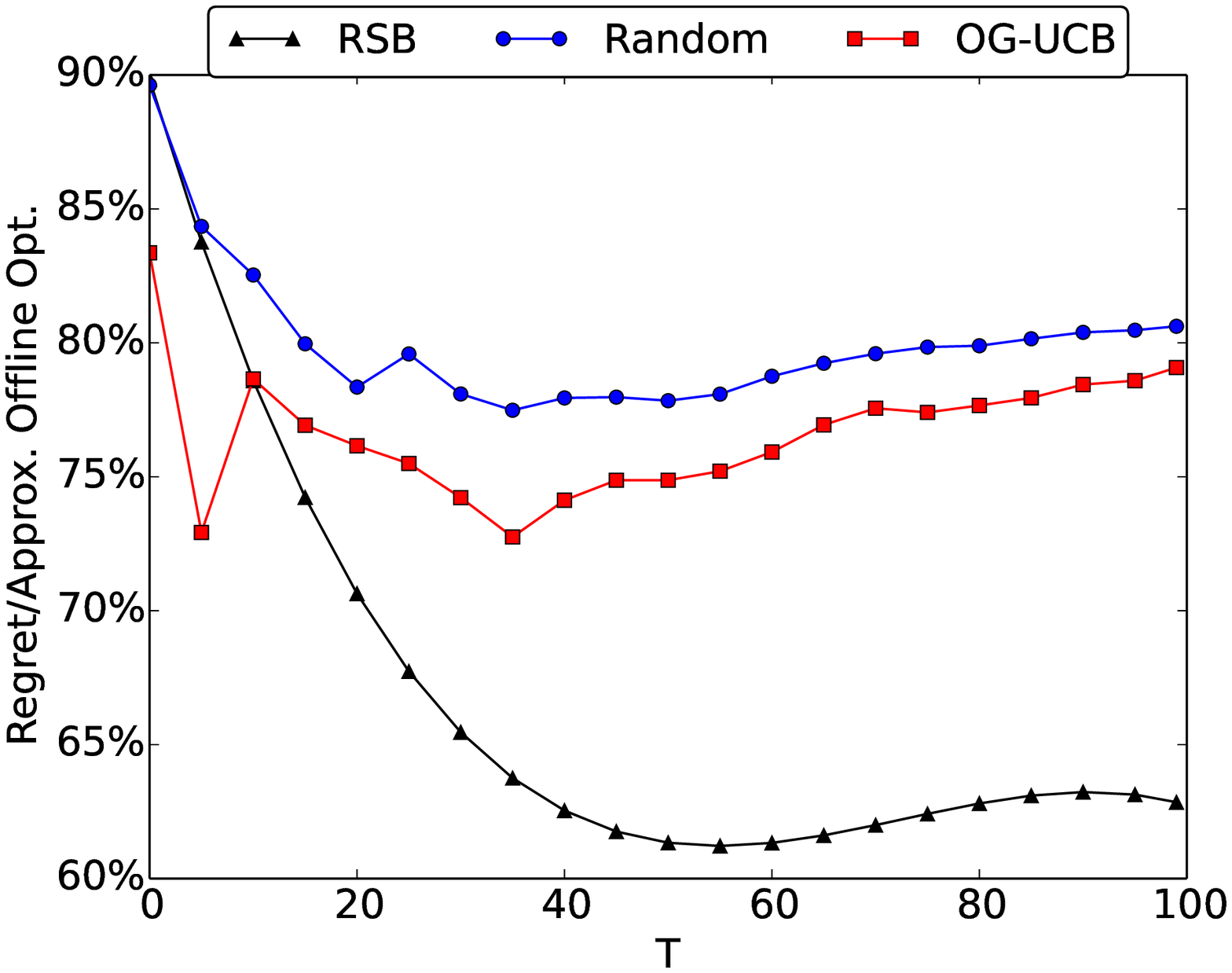}
\caption{Tencent Weibo trace and FR model: $\gamma=0.2$.}
\label{fig4}
\end{minipage}
\hfill
\begin{minipage}[t]{0.32\linewidth}
\centering
\includegraphics[width=\linewidth]{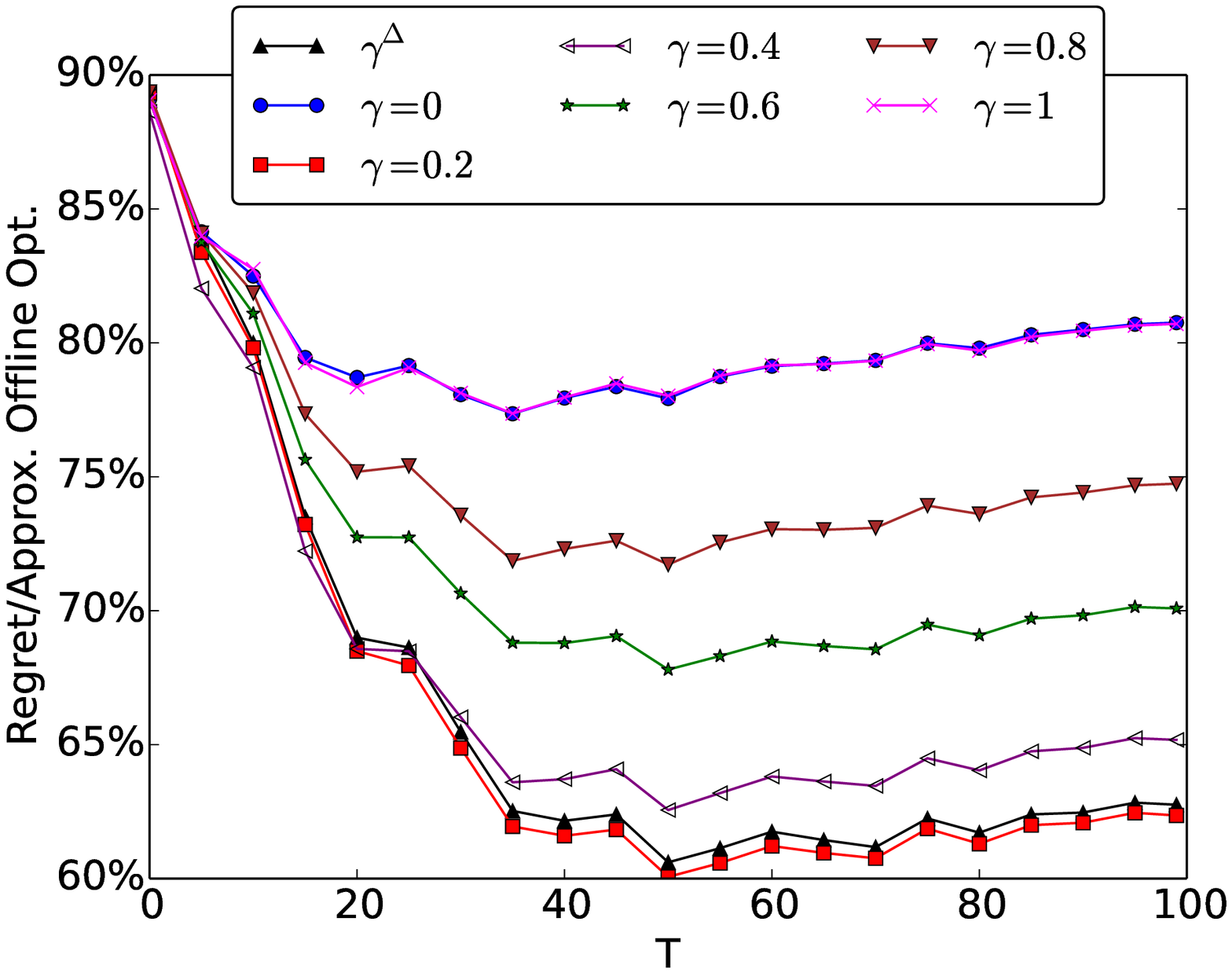}
\caption{Tencent Weibo trace and FR model: different values of $\gamma$.}
\label{fig5}
\end{minipage}
\hfill
\begin{minipage}[t]{0.32\linewidth}
\centering
\includegraphics[width=\linewidth]{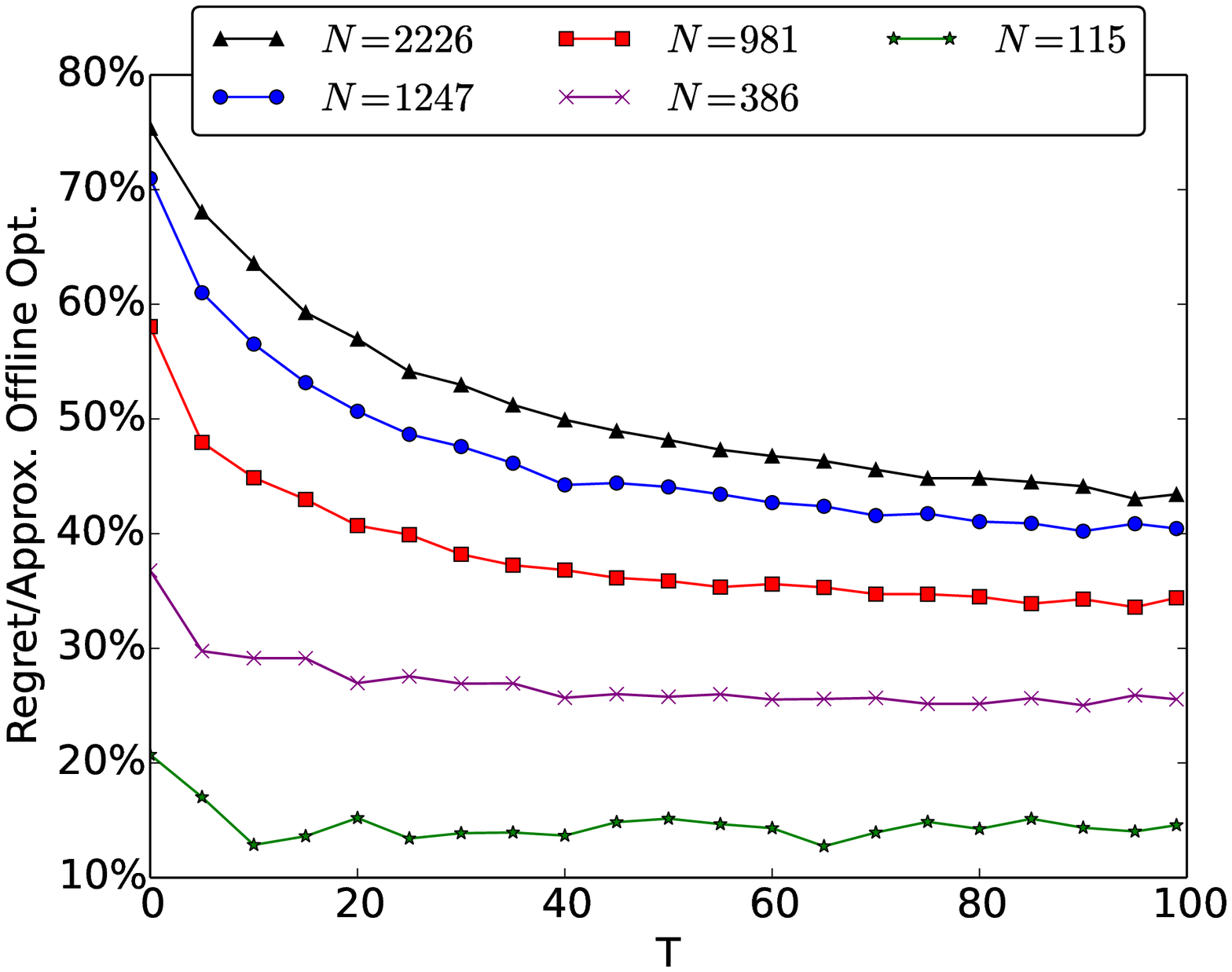}
\caption{Tencent Weibo trace and TR model: different graph sizes $N$.}
\label{fig6}
\end{minipage}
\end{center}
\end{figure*}

Fig.~\ref{fig1}--Fig. \ref{fig4} show the results obtained using synthetic data or Tencent Weibo traces under different time-varying models of influence probabilities. We observe that the regret ratios (and hence information spread) achieved by RSB and the random algorithm are usually similar at the early stages of the system, when RSB has not cumulated much feedback. RSB gets better than the other algorithms (lower regret and hence better spread) after more time stages, validating that RSB can improve with more feedback received from the real system. Besides, OG-UCB performs the worst especially with the ongoing of time, showing that it is only suitable for fixed influence probability distributions and does not work well in cases of time-varying influence probabilities. 
The increase of cumulative regret by RSB with the increase of time stages, if any, is always slower than that of the other algorithms.


In Fig.~\ref{fig5}, we compare the regret ratios of RSB achieved under different values of input parameter $\gamma$, using Tencent Weibo traces under the FR model. From line \ref{linew} of Alg.~\ref{algo2}, we can see $\gamma=0$ represents pure exploitation and $\gamma=1$ indicates pure exploration. Although Theorem \ref{th4} requires $\gamma>0$ for the bound to be meaningful, we can still test the extreme case that $\gamma=0$ when running the algorithm in practice. 
RSB performs worst in these extreme cases. 
$\gamma^{\Delta}=0.18$ is computed following the formula in Theorem \ref{th4} which minimizes the theoretical upper bound. We observe that $\gamma^{\Delta}$ achieves near-lowest regrets in actual execution of our algorithm under practical settings as well.

In Fig.~\ref{fig6}, we evaluate the impact of different graph sizes $N$, by extracting subgraphs of different sizes using Tencent Weibo traces. We observe that the regret is larger in larger networks, but it always improve when the system runs for longer period of time.


\section{Conclusion}\label{sec7}
This paper investigates online influence maximization in dynamic social networks with non-stationary influence probability distributions among participants. We design a randomized algorithm based on multi-armed bandit optimization to guide source selection for information dissemination over multiple time stages, aiming to maximize the overall spread over the system span. The algorithm is simple and neat, relying on carefully designed, continuously updating preferences on seed selection, which exploit real-world feedback from previous decisions, as well as explore new choices. As the first in the literature, the algorithm does not require knowledge of the dynamic social graph topology, nor time-varying influence probabilities, but is able to achieve an upper-bounded weak regret, as compared to an approximate offline optimal reward. Simulations based on both synthetic and real-world datasets further validate that our algorithm is more adaptive to a changing environment than heuristic and stationary bandit algorithms. In addition, our algorithm is also applicable to many other real-world problems such as advertisement placement \cite{chen2013combinatorial}, as long as the reward functions are submodular or there exists an approximate offline algorithm that can achieve an approximation ratio of $(1-\frac{1}{e})$. In future work, we seek to apply similar algorithms to solve the other real-world problems.
\bibliographystyle{IEEEtran}
\bibliography{IEEEabrv,reference}

\begin{appendices}

\section{Proof of Theorem \ref{th2}}
\label{sec:th2}
\begin{IEEEproof}
At any time $t$, given fixed $S_t^{(1:k-1)}$,  there exists a node $a \in S^*$ so that $a \in  \underset{a \in S^*}{\operatorname*{arg\,max}} \sum_{t=1}^T \bar{r}_t^k(a \vert S_t^{(1:k-1)})$. Then $a$ can satisfy the following inequality.
\begin{align}
& \sum_{t=1}^T \bar{r}_t^k(a \vert S_t^{(1:k-1)}) =\sum_{t=1}^T \Big( \bar{r}_t(S_t^{(1:k-1)} \cup \{a\})-\bar{r}_t(S_t^{(1:k-1)})\Big) \nonumber\\
& \geq \frac{1}{K} \Big( \sum_{t=1}^T \bar{r}_t(S^* \cup S_t^{(1:k-1)})-\sum_{t=1}^T \bar{r}_t(S_t^{(1:k-1)})\Big) \label{note1}\\
& \geq \frac{1}{K} \Big( \sum_{t=1}^T \bar{r}_t(S^*)-\sum_{t=1}^T \bar{r}_t(S_t^{(1:k-1)}) \Big) \nonumber
\end{align}

The inequality (\ref{note1}) holds because of pigeonhole principle. As we can select node $a$ with the largest total marginal gain over the whole time period, its marginal reward is equal or larger than the mean value of all nodes belonging to $S^*$.

Note that although $\bar{r}_t^k(a \vert S_t^{(1:k-1)})$ is the expectation taken of random policy's actions, the optimal solution is deterministic thus $\bar{r}_t^k(a \vert S_t^{(1:k-1)})$ reduces to the expectation of random reward only here.

Let $\tilde{a}^k$ be the selected seed with full information fixing $S_t^{(1:k-1)}$, {\em i.e.}, it maximizes the total marginal gain which is equal or larger than $\sum_{t=1}^T f_t^k(a \vert S_t^{(1:k-1)}),\ \forall a \in \mathcal{N}$ under the conditional set $S_t^{(1:k-1)}$. Thus we have
\begin{align*}
 \sum_{t=1}^T \bar{r}_t^k(\tilde{a}^k \vert S_t^{(1:k-1)})
 \geq \frac{1}{K} \Big( \sum_{t=1}^T \bar{r}_t(S^*)-\sum_{t=1}^T \bar{r}_t(S_t^{(1:k-1)}) \Big).
\end{align*}

Define $\Delta(\bar{r}_t^k)=\bar{r}_t(S_t^{(1:k)})-\bar{r}_t(S_t^{(1:k-1)})$. Noting that\\
\begin{align*}
\sum_{t=1}^T \bar{r}_t^k(\tilde{a}^k \vert S_t^{(1:k-1)})-Reg^k(T)=\sum_{t=1}^T \bar{r}_t^k(a^k_t \vert S_t^{(1:k-1)})
\end{align*}
which is the expected value of marginal gain by adding $a^k_t$ to $S_t^{(1:k-1)}$. This implies that
\begin{align*}
\sum_{t=1}^T \Delta(\bar{r}_t^k) =\sum_{t=1}^T \bar{r}_t^k(\tilde{a}^k \vert S_t^{(1:k-1)})-Reg^k(T).
\end{align*}

Then we have
\begin{align*}
\sum_{t=1}^T \Delta(\bar{r}_t^k) \geq \frac{1}{K} \Big( OPT-\sum_{t=1}^T \bar{r}_t(S_t^{(1:k-1)}) \Big)-Reg^k(T).
\end{align*}
\end{IEEEproof}

\section{Proof of Theorem \ref{th3}}
\label{sec:th3}
\begin{IEEEproof}
We prove the following inequality for each position $k$ by induction.
\begin{align} \label{eqOPT}
OPT-\sum_{t=1}^T \bar{r}_t(S_t^{(1:k)}) \leq (1-\frac{1}{K})^k OPT +\sum_{m=1}^k Reg^m(T)
\end{align}

The base case $k=0$ is trivial. In the induction, let
\begin{align*}
Z^k=OPT-\sum_{t=1}^T \bar{r}_t(S_t^{(1:k)})=OPT-\sum_{m=1}^k \sum_{t=1}^T \Delta(\bar{r}_t^m).
\end{align*}

Thus $Z^k=Z^{k-1}-\sum_{t=1}^T \Delta(\bar{r}_t^k)$.

According to Theorem \ref{th2}, we know that
\begin{align*}
\sum_{t=1}^T \Delta(\bar{r}_t^k) \geq \frac{1}{K}Z^{k-1}-Reg^k(T).
\end{align*}

Then we have $Z^k \leq (1-\frac{1}{K}) Z^{k-1}+Reg^k(T)$.

Combining with the induction hypothesis, we can obtain the inequality \ref{eqOPT}. By taking $k=K$ and using $(1-\frac{1}{K})^K<\frac{1}{e}$, we have
\begin{align*}
\sum_{t=1}^T \bar{r}_t(S_t^{(1:K)}) \geq (1-\frac{1}{e}) OPT -\sum_{k=1}^K Reg^k(T).
\end{align*}

Combining with Definition \ref{def3}, the proof is completed.
\end{IEEEproof}

\section{Proof of Theorem \ref{th4}}
\label{sec:th4}
\begin{IEEEproof}
We show the following trivial facts derived from the definitions where $a \vert S_t^{(1:k-1)}$ is the selected arm in Alg.~\ref{algo2}.
\begin{align}
q_t^{n \vert S_t^{(1:k-1)}} \geq w_t^{n \vert S_t^{(1:k-1)}},\ \forall n \in \mathcal{N} \backslash S_t^{(1:k-1)}\nonumber
\end{align}
\begin{align}
& \sum_{n=1}^N w_t^{n \vert S_t^{(1:k-1)}} \hat{r}_t^k(n \vert S_t^{(1:k-1)}) \nonumber\\
& \leq w_t^{a \vert S_t^{(1:k-1)}} \frac{r_t^k(a \vert S_t^{(1:k-1)})}{w_t^{a \vert S_t^{(1:k-1)}}} \nonumber\\
& \leq r_t^k(a \vert S_t^{(1:k-1)}) \label{eqfact2}
\end{align}
\begin{align}
& \sum_{n=1}^N w_t^{n \vert S_t^{(1:k-1)}} (\hat{r}_t^k(n \vert S_t^{(1:k-1)}))^2 \nonumber \\
& \leq w_t^{a \vert S_t^{(1:k-1)}} \frac{r_t^k(a \vert S_t^{(1:k-1)})}{w_t^{a \vert S_t^{(1:k-1)}}} \hat{r}_t^k(a \vert S_t^{(1:k-1)}) \nonumber\\
& \leq D \sum_{n=1}^N \hat{r}_t^k(n \vert S_t^{(1:k-1)}). \label{eqfact3}
\end{align}

We will prove the inequality under any position $k$. In the end we will illustrate that it still holds for the whole $K-$ size seed set.
Under the conditional set $S_{t}^{(1:k-1)}$, $\forall t=1,2,\dots,T$. Let $V_t=\sum_{n=1}^N v_t^{n \vert S_{t}^{(1:k-1)}}$. Then for all actions in Algorithm \ref{algo2}, we have
\begin{align}
& \frac{V_{t+1}}{V_t} = \sum_{n=1}^N \frac{v_{t+1}^{n \vert S_{t+1}^{(1:k-1)}}}{V_t} \nonumber\\
& =\sum_{n=1}^N \frac{v_{t}^{n \vert S_{t}^{(1:k-1)}}}{V_t} \exp{\{\frac{\gamma \hat{r}_t^k(n \vert S_t^{(1:k-1)})}{NC}\}} \nonumber\\
& =\sum_{n=1}^N \frac{w_t^{n \vert S_t^{(1:k-1)}}-\frac{\gamma}{N}}{1-\gamma} \exp{\{\frac{\gamma \hat{r}_t^k(n \vert S_t^{(1:k-1)})}{NC}\}}\nonumber\\
& \leq \sum_{n=1}^N \frac{w_t^{n \vert S_t^{(1:k-1)}}-\frac{\gamma}{N}}{1-\gamma} [1+\frac{\gamma \hat{r}_t^k(n \vert S_t^{(1:k-1)})}{NC}\nonumber\\
& +(e-2)(\frac{\gamma \hat{r}_t^k(n \vert S_t^{(1:k-1)})}{NC})^2] \label{Taylor}\\
& \leq 1+\frac{\frac{\gamma}{NC}}{1-\gamma} r_t^k(a \vert S_t^{(1:k-1)}) \nonumber\\
& + \frac{(e-2)(\frac{\gamma}{NC})^2}{1-\gamma} D \sum_{n=1}^N \hat{r}_t^k(n \vert S_t^{(1:k-1)}).\label{usefact}
\end{align}

The inequality (\ref{Taylor}) uses the fact that $e^x \leq 1+ x+(e-2)x^2$ for $x \leq 1$ and the inequality (\ref{usefact}) is derived by the facts (\ref{eqfact2}) and (\ref{eqfact3}). Using the inequality $1+x \leq e^x$ and taking logarithms, we have
\begin{align*}
& \ln \frac{V_{t+1}}{V_t}\leq \frac{\frac{\gamma}{NC}}{1-\gamma} r_t^k(a \vert S_t^{(1:k-1)})\\
& + \frac{(e-2)(\frac{\gamma}{NC})^2 D}{1-\gamma} \sum_{n=1}^N \hat{r}_t^k(n \vert S_t^{(1:k-1)}). \nonumber
\end{align*}

Let $r^k_{RSB}=\sum_{t=1}^{T} r_t^k(a^k_t \vert S_t^{(1:k-1)})$ and thus $R^k_{RSB}=\mathbb{E}[r^k_{RSB}]$. Then summing over $t$, we can get all reward obtained by the agent over period $1,\dots,T$ for the position $k$ as follows.
\begin{align*}
& \ln \frac{V_{T+1}}{V_1} \leq \frac{\frac{\gamma}{NC}}{1-\gamma} r^k_{RSB} \nonumber \\
& + \frac{(e-2)(\frac{\gamma}{NC})^2 D}{1-\gamma} \sum_{t=1}^{T}\sum_{n=1}^N \hat{r}_t^k(n \vert S_t^{(1:k-1)}) \nonumber
\end{align*}

For any node $n_j \in \mathcal{N}$ whatever the agent selects it, we have
\begin{align*}
\ln \frac{V_{T+1}}{V_{1}} \geq \ln \frac{v_{T+1}^{n_j \vert S_{T+1}^{(1:k-1)}}}{V_{1}}.
\end{align*}

Since $v_{T+1}^{n_j \vert S_{T+1}^{(1:k-1)}}= v_{1}^{n_j \vert S_{1}^{(1:k-1)}} \prod_{t=1}^T \exp{\{\frac{\gamma \hat{r}_{t}^k(n_j \vert S_{t}^{(1:k-1)})}{NC}\}}$ and $v_{1}^{n_j \vert S_{1}^{(1:k-1)}}=1$, we can derive the following inequality.
\begin{align*}
& \ln \frac{V_{T+1}}{V_{1}} \geq \frac{\gamma}{NC} \sum_{t=1}^{T} \hat{r}_{t}^k(n_j \vert S_t^{(1:k-1)})-\ln N
\end{align*}

Based on the inequalities above, we can derive
\begin{align*}
& r^k_{RSB} \geq (1-\gamma)\sum_{t=1}^{T} \hat{r}_{t}^k(n_j \vert S_t^{(1:k-1)}) -\frac{NC \ln N}{\gamma}\\
& -(e-2)\frac{\gamma D}{NC} \sum_{t=1}^{T} \sum_{n=1}^N \hat{r}_{t}^k(n \vert S_t^{(1:k-1)}).
\end{align*}

Then we take the expectation on policy's actions as well as random rewards. Noting that given the choice $a_{1},a_{2},\dots,a_{t-1}$ before, for any node $n_j \in \mathcal{N} \backslash S_t^{(1:k-1)}$, we have
\begin{align}\label{eqEpi}
&\mathbb{E}[\hat{r}_{t}^k(n_j \vert S_t^{(1:k-1)}) \vert a_{1},a_{2},\dots,a_{t-1}]\nonumber\\
&=\mathbb{E}[q_t^{n \vert S_t^{(1:k-1)}} \cdot \frac{r_t^k(n_j \vert S_t^{(1:k-1)})}{q_t^{n \vert S_t^{(1:k-1)}}}+(1-q_t^{n \vert S_t^{(1:k-1)}})\cdot 0] \nonumber \\
&=\bar{r}_t^k(n_j \vert S_t^{(1:k-1)}).
\end{align}

For any node $n_j \in  S_t^{(1:k-1)}$, if we play this node again, the realization of marginal gain might not be zero. But if we consider the expected marginal reward, it must be zero. Since we do not allow the agent to select  the same node as a seed twice in each time stage in Algorithm \ref{algo2}, we have $\hat{r}_{t}^k(n_j \vert S_t^{(1:k-1)})=0$ and $\mathbb{E}[\hat{r}_{t}^k(n_j \vert S_t^{(1:k-1)}) \vert a_{1},a_{2},\dots,a_{t-1}]=0$. Note that $\bar{r}_t^k(n_j \vert S_t^{(1:k-1)})=0$, the equation (\ref{eqEpi}) still holds.

Then we can get
\begin{align*}
& R^k_{RSB} \geq (1-\gamma)\sum_{t=1}^{T} \bar{r}_{t}^k(n_j \vert S_t^{(1:k-1)}) -\frac{NC \ln N}{\gamma}\\
& -(e-2)\frac{\gamma D}{NC} \sum_{t=1}^{T} \sum_{n=1}^N \bar{r}_{t}^k(n \vert S_t^{(1:k-1)}).
\end{align*}

Since over the whole period, the expected marginal gain of any node $n \in \mathcal{N}$ is no larger than that of the best seed, which is $R^k_{\max}$, it is apparent $\sum_{t=1}^{T} \sum_{n=1}^N \bar{r}_{t}^k(n \vert S_t^{(1:k-1)}) \leq NR^k_{\max}$. Since node $n_j$ is chosen arbitrarily, we can choose it as the best seed $\tilde{a}^k$ under the conditional set $S_t^{(1:k-1)}$, thus we have $\sum_{t=1}^{T} \bar{r}_{t}^k(\tilde{a}^k \vert S_t^{(1:k-1)})=R^k_{\max}$.  Combined with these two results, we have
\begin{align*}
R^k_{\max}-R^k_{RSB} \leq (1+(e-2)\frac{D}{C}) \gamma R^k_{\max} + \frac{NC\ln N}{\gamma}.
\end{align*}

Note that the expectation above is under any conditional set $S_t^{(1:k-1)},\ \forall t=1,2,\dots,T$, which is related to previous choices for position $1,\dots,k-1$. This is consistent with Definition \ref{position_regret}, that $S_t^{(1:k-1)}$ is decided by Alg.~\ref{algo2}. The expectation in $R^k_{\max}$ is also reduced to randomizing on reward only. 

Since $\sum_{k=1}^K R^k_{\max} \leq gK$, summing up for all positions $1,2,\dots,K$, we can get
\begin{align}
\sum_{k=1}^K R_{\max}^k-\sum_{k=1}^K R^k_{RSB} \leq (1+(e-2)\frac{D}{C}) \gamma g K + \frac{NCK\ln N}{\gamma}. \label{ineqbound}
\end{align}

Taking the first derivative of the right part in inequality (\ref{ineqbound}), we can set $\gamma=\min\{1,\sqrt{\frac{NC\ln N}{(1+(e-2)\frac{D}{C})g}}\}$ to minimize the bound, then in the period $1,\dots,T$
\begin{align}
& \sum_{k=1}^K Reg^k(T) =\sum_{k=1}^K R^k_{\max}-\sum_{k=1}^K R^k_{RSB }\nonumber\\
& \leq 2K\sqrt{1+(e-2)\frac{D}{C}} \sqrt{gCN \ln N}.\label{eqbound}
\end{align}

Note that if $\gamma=1$, we have $\sqrt{NC\ln N} \geq \sqrt{(1+(e-2)\frac{D}{C})g}$. The bound (\ref{eqbound}) is larger than the maximal reward $Kg$, then it must holds.
\end{IEEEproof}

\section{Proof of Corollary \ref{coro1}}
\label{sec:coro1}
	
\begin{IEEEproof}
It is apparent that the expected number of activated nodes from a seed can not exceed $C$, the size of the largest connected component in the social graph. Then the overall reward achieved over the entire system span can not exceed $CT$. Thus we can set $g=CT$. According to Theorems \ref{th3} and \ref{th4}, we have
\begin{align*}
& Reg_G(T) \leq \sum_{k=1}^K Reg^k(T)\\
& \leq 2K \sqrt{1+(e-2)\frac{D}{C}} \sqrt{DCTN \ln N}.
\end{align*}

Then we have $Reg_G(T)$ is bounded by $O(\sqrt{TN\ln N})$. 
\end{IEEEproof}

\end{appendices}

\end{document}